\documentclass[showkeys]{revtex4-1}

\usepackage{amsfonts}
\usepackage{amsmath}
\usepackage{bbm}
\usepackage[latin2]{inputenc}
\usepackage{graphicx}

\def\iB{\mathcal B}

\def\iC{\mathcal C}

\def\i1{\mathcal 1}

\def\id{\mathrm Id}
\def\1{\mathrm 1}
\def\bbC{{\mathbb C}}

\def\bbN{{\mathbb N}}
\def\>{\rangle}
\def\<{\langle}
\def\Tr{\mathrm Tr}
\def\eps{\varepsilon}
\def\id{\mathbbm{1}}
\def\x{\times}
\def\ox{\otimes}
\begin{document}
\title{\textbf{Simulating quantum systems on the Bethe lattice by translationally invariant infinite-Tree Tensor Network }}
\author{\'Ad\'am Nagy}

\affiliation{Technical University of Budapest
\\
Department of Mathematical Analysis
H-1521 Budapest, POB 91, Hungary
}
\email{turorudi2[at]index.hu, Tel: +36 70 3 441141}

\begin{abstract}
We construct an algorithm to simulate imaginary time evolution of  translationally invariant spin systems with local interactions on an infinite, symmetric tree. We describe the state by symmetric infinite-Tree Tensor Network (iTTN) and use translation-invariant operators for the updates at each time step. The contraction of this tree tensor network can be computed efficiently by recursion without approximations and one can then truncate all the iTTN tensors at the same time. The translational symmetry is preserved at each time step that makes the algorithm very well conditioned and stable. The computational cost scales like $O(D^{q+1})$ with the bond dimension $D$ and coordination number $q$, much favourable than that of the iTEBD on trees [D. Nagaj et al. Phys. Rev. B \textbf{77}, 214431 (2008)].
Studying the transverse-field Ising model on the Bethe lattice, the numerics indicate a ferromagnetic-paramagnetic phase transition, with a finite  correlation length  even at the transition point. 

\end{abstract}

\maketitle

\section{Introduction}

Since S. White's Density Matrix Renormalization Group method \cite{dmrg1}, \cite{dmrg2} was established, many developments have been made in order to simulate interacting spin systems  as they offer an efficient way to study the properties of their ground states and are applicable to fermionic systems as well, see  \cite{review} for a review. The matrix product states description gives us an accurate approximation of the ground states of gapped Hamiltonians on a spin chain, but it can also be applied to detect phase transitions. Various numerical methods have been proposed to carry out these simulations and have achieved decent success.
Their higher dimensional generalizations also appeared  the PEPS \cite{review}, \cite{peps} and its extension to infinite lattices, the iPEPS \cite{roman} and \cite{ipepscorner}, tree tensor networks \cite{ttns0}, \cite{ttns}, MERA  \cite{mera} and other approaches are promising, but their extensive use is hampered by their large (but nevertheless polynomial) computational costs.

In one branch of these methods, that uses MPS description in one spatial dimension and more generally tensor networks in higher dimensions, one describes the state of a spin system in a lattice by placing tensors to each vertex. These tensors has one `physical' index corresponding to the local spin basis vectors, and `virtual indices' for each bond coming out from the vertex. The dimension of this index, the `bond dimension' is a parameter. The quantum mechanical amplitude of a given spin configuration is obtained by fixing the physical indices and contracting all the virtual ones. 
In one dimension, the computation of the expectation of a local operator $O$ i.e. $\<\psi|O|\psi\>$  can be interpreted as the contraction of the chain since we have to sum over all the virtual indices. This can be computed exactly for a fixed bond dimension in 1D, but for higher dimensional lattices this does not hold any more. To see this, consider a subset of the lattice and contract all the indices belonging to the `inner' bonds. Clearly the resulting tensor has a number of indices proportional to the length of the boundary of the subset. Several approximative techniques have been developed (see e.g.   \cite{review}, \cite{terg}) to overcome this and carry out the contraction in polynomial time.  

In this paper, we propose an algorithm to simulate imaginary time evolution of a  translation-invariant  Hamiltonian with local interactions on the Bethe lattice. We essentially generalize the matrix product operator (MPO, \cite{pirvu}) description for higher dimensional lattices and use the fact that a translationally invariant infinite tree tensor network on the Bethe lattice can be contracted efficiently by recursion. This follows directly from the geometry of the tree and is essentially the generalization of the power method for finding the leading eigenvector of a matrix, but instead of the matrix we have a tensor with $q$ indices.   This allows us to update all the iTTN tensors at once in every time step. Thus, all the symmetry is preserved during the whole evolution which makes the algorithm numerically very well conditioned and stable.

Besides  its symmetry preserving and numerical stability, the biggest improvement over the previously established iTEBD method for trees \cite{farhi} is the considerably lower computational cost. In \cite{farhi}, this scaled like $O(D^8)$ for a Bethe lattice with coordination number $q=3$, whereas for the present approach it is $O(D^4)$.

Then we test our new method by simulating the ground state of the transverse field Ising model defined on the infinite tree.
Plotting the transverse and longitudinal magnetizations we find a phase diagram qualitatively similar to that of the Ising model on a spin chain with distinct ferromagnetic and paramagnetic phases. The energy and its first derivative with respect to the external field are continuous, while the second second derivative has a jump. 
Surprisingly, the numerics indicate a finite correlation length for any external field, a striking difference from the one-dimensional counterpart. 
Remarkably, we find a very good agreement with previous findings by the iTEBD method \cite{farhi} and path integral Monte Carlo \cite{montecarlo}.

\section{Contraction of Tree Tensor Networks}

\subsection{Tensors}
Consider a tensor $A$. The number of indices is sometimes called the order or degree of $A$. Although misleading, it is sometimes even called the rank, for example $A_{\alpha\beta\gamma}$ is a 3rd order tensor (or sometimes a 3-way tensor), a 2nd order tensor is a matrix, a first order tensor is a vector.

We will say that $A$ is fully symmetric, if $A_{\alpha_1\alpha_2..\alpha_n}=A_{\alpha_{\pi(1)}\alpha_{\pi(2)}...\alpha_{\pi(n)}}$ for any permutation $\pi$ of the set $\{1,2...n\}$. 

Multiplications of tensors make sense if we specify the indices to be contracted. For instance we can define the j-mode multiplication of the tensor $A\in\bbC^{ \{I_1\x ...I_j...\x I_n  \} }$ by the vector $v\in\bbC^{I_j }$ producing a tensor 
$A\x_j v\in\bbC^{ \{I_1\x..I_{j-1},I_{j+1}..\x I_n  \} }$ as follows
$$
(A\x_j v)_{\alpha_1...\alpha_{j-1},\alpha_{j+1}...\alpha_n}:=\sum_{\alpha_j}A_{\alpha_1...\alpha_j...\alpha_n}v_{\alpha_j}
$$ 
where we also introduced the symbol $\x_j$ as in \cite{hosvd}.
Similarly one can multiply $A$ by the matrix $M\in\bbC^{I_j\x J_j}$, as 
$$
(A\x_j M)_{\alpha_1...\alpha_j...\alpha_n}:=\sum_{\alpha'_j}A_{\alpha_1...\alpha'_j...\alpha_n}M_{\alpha_j\alpha'_j}.
$$
We will also make use of the outer product, i.e. one can make a 3rd order tensor $A$ out of vectors $u,v,w$ as
$$
A_{\alpha\beta\gamma}=(u\circ v\circ w)_{\alpha\beta\gamma}:=u_{\alpha}v_{\beta}w_{\gamma}.
$$
\subsection{Contraction, expectation values and correlations}
\begin{figure}[h]

\includegraphics[width=7cm, height=7cm]{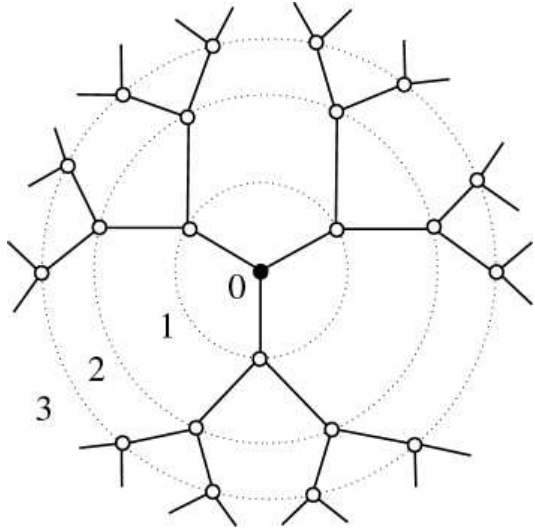}

\caption{Three layers of the Bethe lattice with coordination number $q=3$}\label{fig:1}

\end{figure}

Consider the Bethe lattice (Fig. \ref{fig:1}) with coordination number $q$ and place fully symmetric TTN tensors $A^s$ to each node. This describes the  translation-invariant state $\psi$; from each node, the tree looks the same.

Observe that the computation of the norm square $\<\psi|\psi\>$ on trees essentially reduces to the computation of the leading vector $r$ of the transfer matrix $E=\sum_s A^s\ox\overline{A^s}$, and it can be done efficiently by recursion.
This statement is well known and easy to see in 1D, but similarly simple also for trees. 
Without loss of generality take $q=3$ and use tensor notations for simplicity. Begin with the 3-way, fully symmetric, $d\x d\x d$ tensor $E$ and a random vector $r^{(0)}$ of  dimension $d$.
For $n=0,1,2...$ define the series 
\begin{equation}
r^{(n+1)}=E\x_1 r^{(n)}\x_2 r^{(n)}.
\label{recursion}
\end{equation}
The normalized series  converges to the so-called leading vector of $E$, i.e. $r^{(n)}/\| r^{(n)} \|\to r$, given  $\<r|r_0\>\neq 0$
so the leading vector $r$ is a fix point of the recursion (\ref{recursion}) 
$$
r=E\x_1 r\x_2 r.
$$
Notice that this is just the extension of the power iteration, which is originally an algorithm for finding the dominant eigenvector of a matrix.
 Its generalization for the case where $E$ is not symmetric and has different leading vectors for each mode is straightforward.
Notice also, that for a generic 3rd order tensor $T$, its leading vectors $u,v,w$ give the best rank-one decomposition \cite{tensordecomp}, i.e. minimize
$$
||T-\lambda u\circ v\circ w||
$$
and $\lambda=T\x_1 u\x_2 v\x_3 w$, assuming $||u||=||v||=||w||=1$. 

Using the leading vector $r$ one gets
$$
\<\psi|\psi\>=\<\overline{r}|r\>=\sum_{\alpha}r_{\alpha}^2.
$$ 
The expectation of a local operator $O$ in the state $\psi$ reads
$$
\<O\>=\frac{\<\overline{r}|E_O|r\>}{\<\overline{r}|E|r\>}
$$
with $E_O=\sum_{s,s'} \<s|O|s'\> A^s\ox \overline{A^{s'}}$,
and the correlation between two sites of distance $n$  reduces to the computation of the second largest eigenvalue of the matrix
$$
M=E\x_1 r.
$$
This is in accordance with the findings of \cite{farhi} but written in a much more compact form.
\subsection{Computational costs}

Clearly, the direct computational cost of each iteration step  scales as $O(d^q)$ for arbitrary $q\in\bbN$. 
As for the convergence, we can state the followings.
If the fully symmetric $E$ can be decomposed such that
$$
E=\sum_i\lambda_i v_i\circ v_i\circ v_i
$$
where $\<v_i|v_j\>=0$ for $i\neq j$ and $|\lambda_1|>|\lambda_2|\geq|\lambda_3|\geq...$ then the convergence is surely geometric with ratio $|\lambda_2/\lambda_1|$. This corresponds to the matrix case when the matrix is diagonalizable.
In any case, given a concrete $E$, the convergence should be checked by doing the numerics.
Let us mention that since $|\lambda_2/\lambda_1|$ is related to the correlation length of the system, one can expect an exponentially fast convergence only if the correlation length is finite. Otherwise, if the  correlation length diverges as the function of the bond dimension, the algorithm slows down.

However, we can exploit the symmetries of $E$ to get a much better scaling.
To this end, simply exploit the structure of $E=\sum_s A^s\ox\overline{A^s}$.
Rewrite (\ref{recursion}) explicitly
\begin{equation}
r_{\gamma\gamma'}^{(n+1)}=\sum_{s=0}^1\sum_{\alpha\alpha'\beta\beta'}A_{\alpha\beta\gamma}^s\overline{A^s_{\alpha'\beta'\gamma'}}r_{\alpha\alpha'}^{(n)}v_{\beta\beta'}^{(n)}
\end{equation}
where all the Greek indices run from $ 1$ to $D $. Now break the sum into four parts. First let us sum over $\alpha$, then over $\beta$. When regarding $r_{\alpha\alpha'}$ as a matrix, we will write $R$, i.e. define
the tensor $\iB^s=A^s\x_1 R^{(n)}$, explicitly written
$$
\iB_{\alpha'\beta\gamma}^s:=\sum_{\alpha=1}^DA_{\alpha\beta\gamma}^sr_{\alpha\alpha'}^{(n)}.
$$
This requires $O(D^4)$ steps because we fix $s,\alpha',\beta,\gamma$ and sum over $\alpha$.  
In the same way perform 
$$
\iC_{\alpha'\beta'\gamma}^s:=\sum_{\beta=1}^D\iB_{\alpha'\beta\gamma}^sr_{\beta\beta'}^{(n)}.
$$
Finally we arrive at

$$
r_{\gamma\gamma'}^{(n+1)}=\sum_{s=0}^1\sum_{\alpha'\beta'=1}^D\iC_{\alpha'\beta'\gamma}^s\overline{A_{\alpha'\beta'\gamma'}^s}.
$$
again at a cost of $O(D^4)$.
Performing the recursion as described above requires $O(D^4)$ computational steps.
Its easy to see that the idea works for arbitrary $q$ and its cost scales like $O(D^{q+1})$.

Note that, the above argument still holds for the computation of the expectation value $\<O\>$ .

In order to compute the correlation length one has to find the second largest eigenvalue of the $D^2\x D^2$ matrix $M$ at a cost of $O(D^4)$ steps.

\subsection{Tensor updates}

Let us consider a time evolution induced by a translationally invariant Hamiltonian on the tree lattice. If one starts with a translation-invariant state, then the symmetry of the Hamiltonian guarantees that it is never broken in time. 
Assume we have a TTN description with bond dimension $D$ for this translationa-invariant state $\psi$. Obviously each site has the same TTN tensors $A^s$.
Now we update these tensors in every time step and obtain the updated state $\psi'$ with new TTN tensors and bond dimension $2D$. Then we need to truncate them in an optimal way, i.e. we should find TTN tensors $\widetilde{A^s}$ with bond dimension $D$ describing a state $\widetilde{\psi}$ which minimizes the distance from $\psi'$, i.e. for which
$$
\|\psi'-\widetilde{\psi}\|
$$
is minimal.

In order to carry out the optimal cut express the norm square by the leading vector $r'$ of the updated $E'$ as
$$
\<\psi'|\psi'\>=E'\x_1r'\x_2r'\x_3r'=r'\x r'=\sum_{\gamma\gamma'=1}^{2D}r'_{\gamma\gamma'}r'_{\gamma\gamma'}=\<\overline{r'}|r'\>.
$$
One can look at $r'_{\gamma\gamma'}$ as a symmetric  $2D\x 2D$, complex matrix which we will denote by $R'=R'^\dagger$.
Perform its singular value decomposition: $R'=U\Lambda U^\dagger$ and introduce the $D\x 2D$ matrix $\widetilde{U}$ consisting of the first $D$ rows of $U$ and $\widetilde{\Lambda}$ is a $D\x D$ diagonal matrix  made up of the $D$ largest singular values. Define the self adjoint projection $P=\widetilde{U}^\dagger\widetilde{U}$, indeed
$P=P^\dagger=P^2$ since $\widetilde{U}\widetilde{U}^\dagger=\id_D$. We can project to the subspace of the highest $D$ singular values as $A^s\x_1P\x_2P\x_3P$, 
implying that the appropriate cut is then given by the transformation 
$$
A^s\longrightarrow \widetilde{A^s}=A^s\x_1\widetilde{U}\x_2\widetilde{U}\x_3\widetilde{U}.
$$
These new tensors will have the right size $D\x D\x D$ and  describe the state $\widetilde{\psi}$.
This is nothing else than the extension of the 1D algorithm to the tree geometry. Recall that in one dimension the truncation step looks like \cite{pirvu}
$$
A^s\longrightarrow \widetilde{A^s}=\widetilde{U} A^s\widetilde{U}^\dagger= A^s\x_1\widetilde{U}\x_2\widetilde{U}.
$$
As always, the lower the discarded weight relatively, the higher the accuracy of the approximation. If the former is not small enough, one should increase the bond dimension.
\vspace{2mm}

\textbf{Translation-invariant iTEBD on the Bethe lattice}

\vspace{2 mm}

We also show that the one dimensional iTEBD algorithm \cite{itebd} which uses the so-called canonical representation of infinite-MPS (iMPS, see e.g. the review \cite{review}) can also be extended to the Bethe lattice. Note that here one updates all the tensors at once therefore  translation invariance is kept at each step, given that the infinite-MPO (iMPO) and the starting state have been so. Hence this update method differs from the one in \cite{farhi} where instead of iMPO, the interaction between only two adjacent sites has been considered at each step, leading to the breaking of the translational symmetry during the process. 

\textbf{Canonical iTTN}

This representation contains the tensors $\Gamma$, placing to the sites and the explicit Schmidt coefficients $\lambda$ (sitting on the bonds) that one gets if cutting the tree between two neighbouring sites \cite{farhi}
$$
|\psi\>=\prod_{k\in \text{bonds}}\sum_{\alpha_k} \lambda_{\alpha_k}^{(k)} \prod_{i\in\text{sites}} \sum_{s_i,\alpha_l...}\Gamma^{(i),s_i}_{\alpha_l,\alpha_m,\alpha_n}|..\>|s_i\>|..\>
$$
each index $\alpha_l$ appears in two $\Gamma$ tensors and one $\lambda$.
The normalization conditions for the canonical iTTN on the Bethe lattice look like
\begin{equation}
\sum_{\alpha}\lambda_\alpha^2=1,
\end{equation}
\begin{equation}
\sum_{s}\sum_{\alpha_k}\sum_{\alpha_l} \overline{\Gamma^{(i),s_i}_{\alpha_k,\alpha_l,\alpha_m'}}\lambda_{\alpha_k}^{(k)2}\lambda_{\alpha_l}^{(l)2}
\Gamma^{(i),s_i}_{\alpha_k,\alpha_l,\alpha_m}=\delta_{\alpha_m,\alpha_{m'}}
\label{normcond}
\end{equation}
\begin{figure}

\includegraphics[width=9cm, height=6cm]{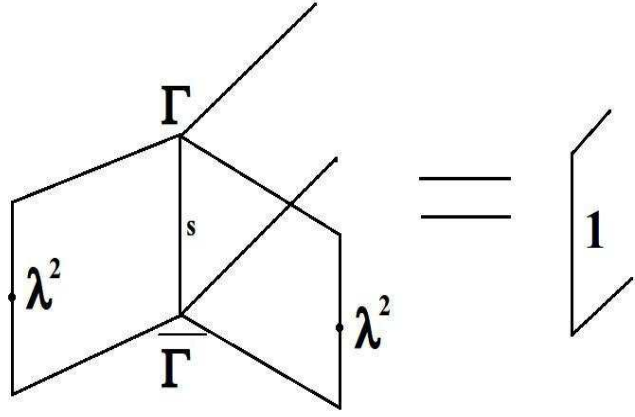}

\caption{The norm condition (\ref{normcond}) in diagram}\label{fig:2}

\end{figure}
this latter relation becomes much more unambiguous if represented in diagram, see Fig.\ref{fig:2}.
There are two conditions like this if $q=3$, interchanging the legs of the tensor $\Gamma$, i.e. we have two layers due to the tensor product, and we leave two open legs (indices) in one direction and close the other legs by placing $\lambda^2$ and sum over the corresponding bonds. 
The directional symmetry manifests itself in a fully symmetric $\Gamma^s$.
Now applying the iMPO to this iTTN results in higher bond dimension and the violation of these norm conditions: The new leading vector will be different from the identity. According to the procedure in \cite{itebd}, we have to bring this back to the canonical form (orthogonalization) and then truncate all bond indices by retaining the $D$ highest Schmidt coefficients.
We can naturally extend this by performing the same operations on each leg as in the one dimensional case, noting that we can define the dominant eigenvector by (\ref{recursion}) and in case of directional symmetry, this will be the same for all legs. After computing this leading vector $R$ (which is interpreted as a matrix because of the two layers), we perform the same manipulations as in Fig. 3 in \cite{itebd}, but now we have open legs in one direction and `closed legs' in the other $q-1$ directions.

\section{Imaginary Time Evolution }

Consider the Hamiltonian on the Bethe tree $G = (V, E)$ with coordination number $q$
\begin{equation}
H=-\sum_{i,j\in V\ i\sim j} J\sigma_i^x\sigma_j^x-\sum_{i\in V}h\sigma_i^z
\end{equation}

As shown in \cite{pirvu} one can derive an elegant MPO representation for $e^{-\eps H}$ with $H$ being the Ising Hamiltonian on a one-dimensional infinite chain.
\begin{equation}
\exp(-\eps H)=\sum_{k_1,k_2..}\Tr(C^{k_1}(\eps)C^{k_2}(\eps)...)Z^{k_1}\ox Z^{k_2}...
\label{MPOising}
\end{equation}
with 2 dimensional MPO matrices $C^k(\epsilon)$ given by
\begin{eqnarray*}
C^0(\eps)&=& \sum_i B_iB_i^{T}=\left(\begin{array}{cc}\cosh\eps & 0 \\0 & \sinh\eps \end{array}\right) \\
C^1(\eps)&=& \sum_i B_{i\oplus 1}B_i^{T}=\left(\begin{array}{cc}0 & \sqrt{\cosh\eps\sinh\eps} \\\sqrt{\cosh\eps\sinh\eps} & \ 0 \end{array}\right).
\end{eqnarray*}

And the spin operators are $X^0=e^{\delta\sigma^z}$ , $X^1=\sigma^x$, $\delta=h\Delta t/2$ for the second order Suzuki-Trotter
formula \cite{pirvu}. We would like to determine the local `transfer tensors' $Q^{ss'}_{\alpha\beta\gamma}$ that makes the TTN tensors evolve as
$$
A^s_{\alpha\beta\gamma}\longrightarrow \sum_{s'} A^{s'}_{\alpha\beta\gamma}Q^{s's}_{\alpha'\beta'\gamma'},
$$ 
or using tensor notation
$$
A^s\longrightarrow \sum_{s'} A^{s'}\ox Q^{s's}.
$$
We can express the MPO matrices by the sum of outer product of vectors. First define the vectors
$$
v=\big ( \sqrt{\cosh(J\Delta t)},\ \sqrt{\sinh(J\Delta t)}\big )
$$
and 
$$
u=\big( \exp(h\Delta t/2),\ \exp(-h\Delta t/2)\big ).
$$ 
Then observe that in 1D, the MPO-s has the compact form
$$
C_{\alpha\beta}^k=v_{\alpha}v_{\beta}\id\{k=\alpha+\beta\}
$$
$\alpha,\beta$ corresponding to the $x$ direction, and $\id\{A\}$ being the indicator function of the event $A$.

The translationally invariant local transfer tensors read
$$
Q_{\alpha\beta}^{ss'}=v_{\alpha}v_{\beta}u_{s}u_{s'}\id\{s+s'=\alpha+\beta\}.
$$
analogously, if there are three spatial indices, the MPO-s has the form
$$
C_{\alpha\beta\gamma}^k=v_{\alpha}v_{\beta}v_{\gamma}, \id\{ \alpha+\beta+\gamma=k\}
$$

and the translationally invariant local transfer tensors read
$$
Q_{\alpha\beta\gamma}^{ss'}=v_{\alpha}v_{\beta}v_{\gamma}u_{s}u_{s'}\id\{s+s'=\alpha+\beta+\gamma\}.
$$
this can be written as a sum of two rank-one tensors as can be verified 
$$
Q=u\circ u\circ v\circ v\circ v+\tilde{u}\circ \tilde{u}\circ \tilde{v}\circ \tilde{v}\circ \tilde{v}
$$
using tensor notation and $\tilde{u}=\sigma^z u$, $\tilde{v}=\sigma^z v$.

Note that the above derivation for the local interactions like $\sigma_x\ox\sigma_x$ with a transverse-field can be repeated for $\sigma_y\ox\sigma_y$ or $\sigma_z\ox\sigma_z$.

\section{Results and discussion}

\begin{figure}[h!]

\centering

\includegraphics[width=12cm, height=8cm]{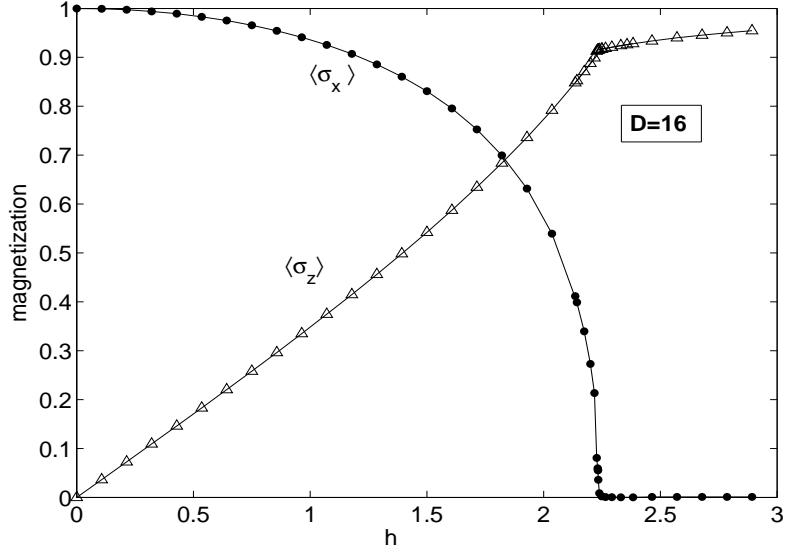}

\caption{Longitudinal and transverse magnetization of the Ising model on the $q=3$ Bethe lattice.}\label{fig:3}

\end{figure}

\begin{figure}[h!]

\centering

\includegraphics[width=12cm, height=8cm]{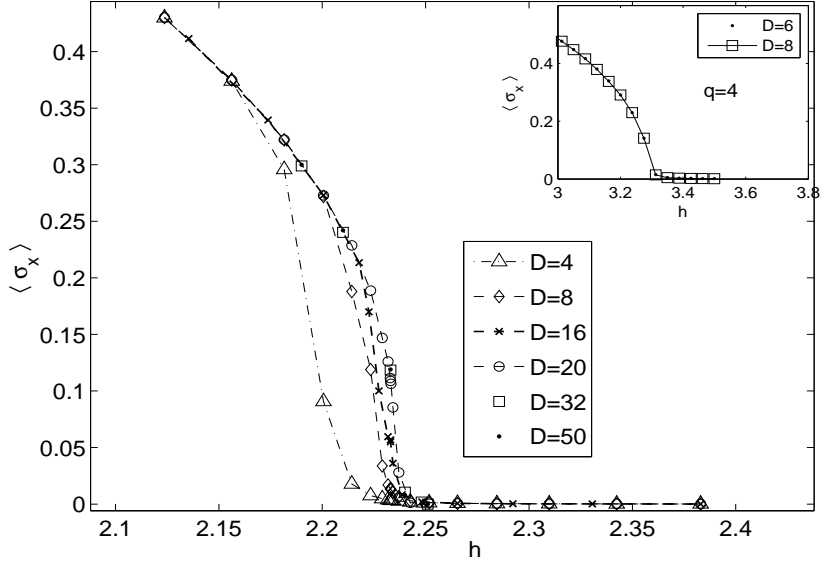}

\caption{Longitudinal magnetization of the Ising model on the $q=3$ and $q=4$ (subplot) Bethe lattice around the transition point. Notice that even for small bond dimensions, the results are barely different.}\label{fig:3}

\end{figure}

We set $J=1$ making interactions ferromagnetic and vary $h$.

\begin{figure}[h!]

\centering

\includegraphics[width=11cm, height=6cm]{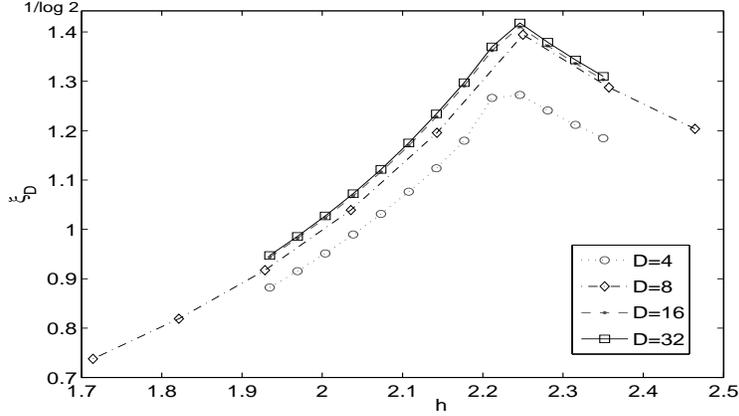}

\caption{Correlation length around the transition point for different bond dimensions, $q=3$}\label{fig:4}

\end{figure}

Using the present algorithm I found very good agreement with the results reported earlier. Namely, the transverse and parallel magnetization  as the function of the magnetic field behave similarly to that of the Ising model on an infinite line, but the critical point is shifted, and estimated to be in the interval $2.23<h/J<2.25$.

One can also check whether the longitudinal magnetization obeys the scaling law 
$$
\langle\sigma_x\rangle\propto (h_c-h)^\beta
$$
in the ferromagnetic phase, near to the transition point. Fitting a line in the log-log plot, we obtain $\beta\approx 0.46$. In \cite{farhi} they reported $\beta=0.41$, while the so-called `cavity method' prediction  $\beta=1/2$ equals the mean-field result, see \cite{montecarlo}.

\begin{figure}[h!]

\includegraphics[width=11cm, height=6cm]{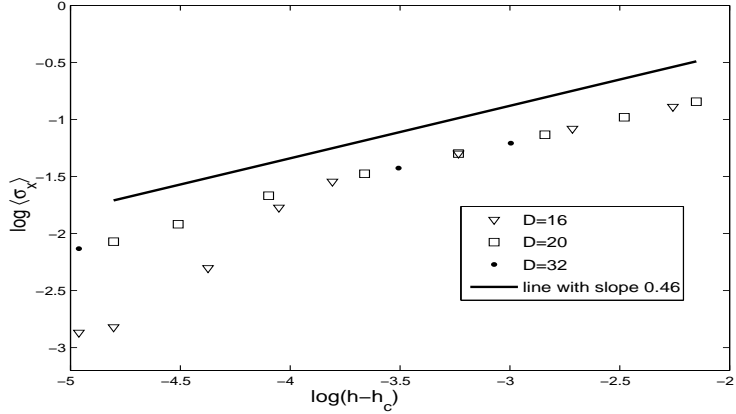}

\caption{Log-log plot of the parallel magnetization vs $h-h_c$}\label{fig:critexp}

\end{figure}

\begin{figure}[h!]

\centering

\includegraphics[width=14cm, height=8cm]{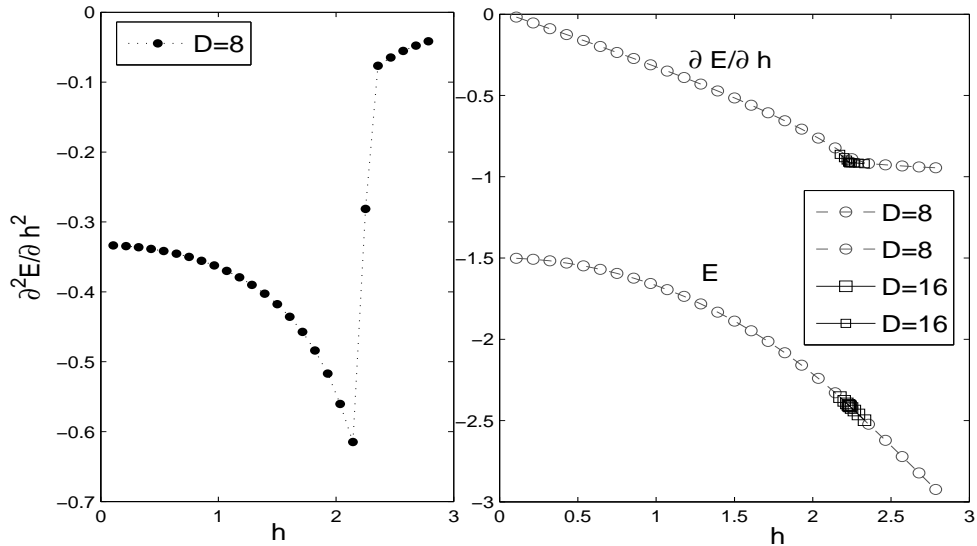}

\caption{Ground state energy and its first and second derivatives with respect to $h$ for  $q=3$.
Notice that the results for $D=8$ and $D=16$ are indistinguishable in the figure.}\label{fig:5}

\end{figure}

Surprisingly, the correlation length $\xi$ at the transition point does not seem to diverge as $D\rightarrow\infty$, since the second largest eigenvalue $\lambda_2$ of $B$ never exceeds $1/2$, in complete agreement with the findings of \cite{farhi}.
This is a strong numerical evidence of a finite correlation length even in a presence of a (supposed) phase transition.
 Also, plotting the first derivative of the energy per site with respect to the magnetic field, we see a non analytic breaking-point at the suspected critical field, while the plot suggest a discontinuous second derivative with a finite jump at the same point, see Fig. \ref{fig:5}.
Qualitatively the ground state energy behaves very similarly to that of the  exactly solvable quantum Curie-Weiss model \cite{montecarlo}.

Note that in \cite{farhi} they used a different parametrization of the Hamiltonian with the parameter $s$ for which $h/J=3(1-s)/s$.

I also considered the Bethe lattice with coordination number $q=4$ and found similar phase diagrams, but with transition point at 
$h/J\approx 3.3$ and  the results suggest that $\lambda_2\leq 1/3$. 

\vspace*{2pt}

\textbf{Previous results and theoretical considerations}

\vspace*{2pt}

In \cite{cavity} the authors extended the cavity method to quantum spin-1/2 models making it  applicable to Bethe lattices, too.
In \cite{montecarlo}, by using Monte Carlo simulations, it is applied to the present model. They computed the magnetization for different temperatures, and their zero temperature extrapolations (e.g. for the ground state) agree very well with the present findings and  with \cite{farhi}. 

We would like to draw the reader's attention to the last section of \cite{farhi}, where the two point correlation is calculated.
Given a translation-invariant Hamiltonian with nearest neighbour interactions on the Bethe lattice, it is shown that the two point correlations always fall off exponentially, and the correlation length is upper bounded by $1/log(q-1)$, confirming our numerical findings. 
According to this brief argument, it is a model-independent and computational method independent consequence of assuming that the translational independent ground state is the stable limit of a sequence of ground states as the size of the tree grows.

Notice that the argument also holds for the classical Ising model defined on the Bethe lattice which is exactly solvable and exhibits phase transition \cite{baxter}.

Note also, that in \cite{montecarlo}, the correlation length was not computed, but (mistakenly) supposed to be infinite at the critical point.

\textbf{Errors}

Concerning the errors one can distinguish two sources: using the Trotter-Suzuki formula and the truncation.
The former has the order of $N(\Delta t)^3=t^3/N^2$, as the second order Trotter formula is used, i.e.
$e^{-tH}\approx \Big ( e^{-\frac{t}{2N}H_z}e^{-\frac{t}{N}H_x}e^{-\frac{t}{2N}H_z}\Big )^N$.
So we need $t\rightarrow\infty$ to obtain the ground state properties and $N\rightarrow\infty$ together with $\Delta t=t/N\rightarrow 0$ and $N(\Delta t)^3=t^3/N^2\rightarrow 0$ to make the Trotter error go to zero. This can be achieved e.g. by setting $N(t)=ct^2$  for some constant $c$ and then increasing $t$ until the results do not change significantly.

\begin{figure}[h!]

\centering

\includegraphics[width=14cm, height=8cm]{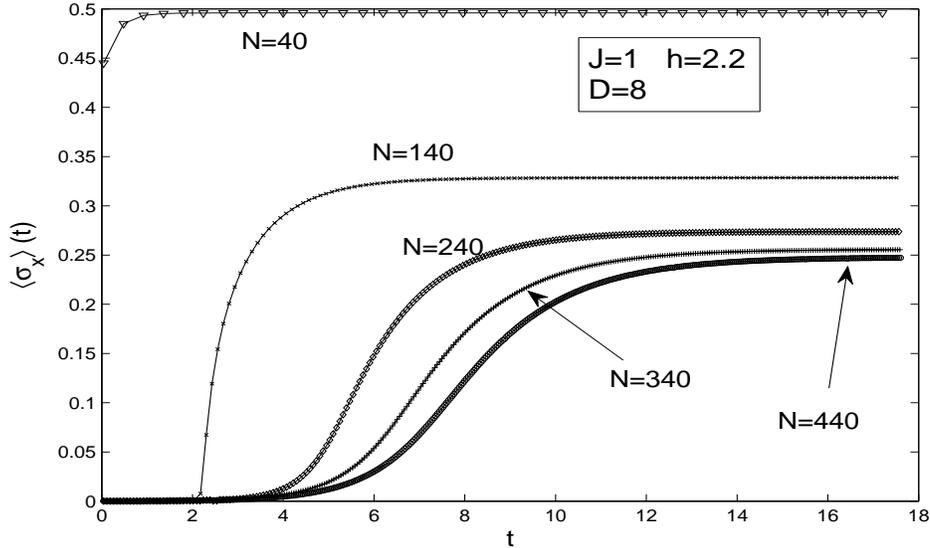}

\caption{Typical behaviour of a computed physical quantity during the time evolution. The maximum time $T$ is fixed, and the time steps $N$ are increased linearly. Starting from the same initial state,  we get different curves due to the Trotter errors.}\label{fig:8}

\end{figure}

In Fig. \ref{fig:8} we plot the longitudinal magnetization $m_x=\langle \sigma_x \rangle$ computed as a function of time.
Here we fix the maximum time $T$ and run the algorithm five times with different number of time steps $N$.
One can see that higher the $N$ the slower the convergence to the limiting value $m_x(T,N)$, but as a function of $N$ (or $\Delta t=T/N$) the convergence to the limit $m_x(T)=\lim_{N\rightarrow\infty} m_x(T,N)$ is very fast. 
Then increasing the time $m_x(T)\rightarrow m_x$, in the present case we can choose $N(T)=2T^2$. In general, if the system is gapless, this convergence is exponential in $T$.

Note that one has to be cautious about the convergence of different quantities, usually the convergence of the energy is much faster than that of the magnetization, so it is not sufficient to check only the convergence of the energy.

\section{Summary}
In this paper we have derived an algorithm to simulate the imaginary time evolution induced by certain translationally invariant local Hamiltonians on the Bethe lattice.
During the time evolution we have used a translationally invariant iTTN description of the evolving state and have given the optimal updates which retain the symmetry at every step. 
The presented method has much lower computational costs than that of the slightly modificated version of the iTEBD method for trees.
reported in \cite{farhi}.
That scaled like $O(D^8)$, and the  translational symmetry was broken during the algorithm.
Investigated specifically the Ising model on trees, we have found  very good agreement with previous numerical results, namely a  phase transition between a ferromagnetic and a paramagnetic phase but with finite correlation length even at the transition point.
At this critical point, the second derivative of the energy with respect to the magnetic field has a finite jump, while the energy and its first derivative is continuous.
Considering the results for $q=2,3,4$ suggests that increasing the coordination number $q$, at fixed $J$, the critical magnetic field increases, and the maximal correlation length decreases.

\end{document}